\documentclass[onecolumn,secnumarabic,amssymb,nobibnotes,showpacs, aps, prb]{revtex4}
\usepackage{epsfig}
\usepackage{graphicx}
\usepackage{dcolumn}
\usepackage{epsfig,rotating,graphicx,color}

\begin{document}

\bibliographystyle{apsrev}
\preprint{code number:}
\title{Role of interfaces in the biased composition of TbFe(Co) thin films }

\author{E. Haltz$^1$, R. Weil$^{1}$, J. Sampaio$^1$, A. Pointillon$^1$, O. Rousseau$^1$,  K. March$^1$, N. Brun$^1$, Z. Li$^1$, E. Briand$^2$, C. Bachelet$^3$, Y. Dumont$^4$ and A. Mougin$^1$ } \email{alexandra.mougin@u-psud.fr} \affiliation{$^1$Laboratoire de Physique des Solides, CNRS, Univ. Paris-Sud, Universit\'e Paris-Saclay,  B\^at.~510,  91405 Orsay, France} \affiliation{$^2$Institut des NanoSciences de Paris (UMR7588), 4 place Jussieu, 75252 Paris Cedex 05, France } \affiliation{$^3$Centre de Sciences Nucl\'eaires et des Sciences de la Mati\`ere, CNRS, Univ. Paris-Sud, Universit\'e Paris-Saclay,  B\^at.~108,  91405 Orsay, France}\affiliation{$^4$Groupe d'Etude de la Mati\`ere Condens\'ee (GEMaC), UMR Universit\'e de Versailles St Quentin en Y. and CNRS, Universit\'e Paris-Saclay, Versailles, France}
\date{\today}
%
%
\begin{abstract}
Ferrimagnetic TbFe or TbFeCo amorphous alloy thin films have been grown by co-evaporation in ultra-high vacuum.  They exhibit an out-of-plane magnetic anisotropy up to their Curie temperature with a nucleation and propagation reversal mechanism suitable for current induced domain wall motion. Rutherford back scattering experiments confirmed a fine control of the Tb depth-integrated composition within the evaporation process. However, a large set of experimental techniques were used to evidence an interface related contribution in such thin films as compared to much thicker samples. In particular, scanning transmission electron microscopy experiments evidence a depth dependent composition and perturbed top and bottom interfaces with preferential oxidation and diffusion of terbium. Despite of that, amorphous and homogeneous alloy film remains in a bulk-like part. The composition of that bulk-like part of the magnetic layer, labeled as effective composition, is biased when compared with the depth-integrated composition. The magnetic properties of the film are mostly dictated by this effective composition, which we show changes with different top and bottom interfaces.  
\end{abstract}

\maketitle

\section{Introduction}

Recent progress have highlighted the interest to spintronics of systems with small magnetization, such as ferrimagnets, antiferromagnets or synthetic antiferromagnetic stacks~\cite{Review_AF_2018}. This has driven the research in Spin Transfer Torque towards news materials and stacks.

Ferrimagnetic alloys  built from the combination of two types of magnetic elements Rare Earths (RE) and Transition Metals (TM) with antiferromagnetically coupled sublattices display magnetic properties (as magnetization and angular momenta) tunable with temperature and composition~\cite{hansen-JAP-66-756-1989,Kirilyuk2013}. Interestingly, it has been shown in the 70s that relatively thick amorphous rare earth transition metals compounds films could have a strong perpendicular anisotropy~\cite{Chaudhari1973} which made them very useful for magneto-optical recording~\cite{mansuripur-IEEE-22-33-1986}. Recently,  the interest of such alloys was renewed by the investigations of current induced domain wall motion in ultra-thin films with perpendicular anisotropy. 
In ferrimagnets, low net magnetization ($M_S$) promotes the N\'eel DWs interesting for spin orbit torques DW motion in stacks including heavy metals~\cite{Haazen2013}. This increases the efficiency  of current-induced torques~\cite{Komine_JAP2011_ferri_STT_calculs,Hirata_arXiv_2017,Haltz_sub_2018} whereas the low magnetization decreases the depinning field~\cite{Nishimura2018}. Eventually, at the angular compensation with a non-zero magnetization, field-driven antiferromagnetic spin dynamics is realized in ferrimagnets with a much faster DW propagation~\cite{Yuushou2018,Kim2017}.  RETM ferrimagnets are also appealing materials to promote skyrmions~\cite{Woo2018}. However, all these new perspectives require a fine controlled of the interfacial properties or environment (adjacent heavy metal layers) and are scarly referenced up to day for such ferrimagnetic alloys in reduced dimensions.

The main purpose of this paper is to present the  structural and chemical properties of TbFe or TbFeCo alloys thin films stacks grown by co-evaporation in relation with their magnetic behavior. The influence of the layers surrounding the ferrimagnet is discussed. We show that the effective composition of a bulk-like part is biased due to interfaces and that this effective composition tunes compensation, one of the most sensitive fingerprint of TbFe(Co) alloys thin films stacks.

 \section{Growth and structure of amorphous //TbFe(Co)/Al alloys thin films}

\subsection{UHV co-evaporation protocol} 

TbFe (respectively TbFeCo) alloys have been grown by co-evaporation of Tb and Fe (respectively a fused Fe$_{85}$Co$_{15}$ target) in a dedicated UHV chamber. Its base pressure is about 2.10$^{-10}$ mbar and it's equipped with two electron guns of 7 crucibles each, allowing to develop complex multilayer stacks. For growing ferrimagnetic alloy films, pure Fe (Fe$_{85}$Co$_{15}$) and pure Tb were co-evaporated. The tuning of the alloy composition is a prerequisite to any further study. For that purpose, three independent quartz balances are available in the chamber: one in the vicinity of each electron gun and a removable one that can be positioned at the exact substrate position to perform a precise determination of the geometrical tooling factors between the flux received by the sample and those measured above the electron guns.   During growth, the evaporation rates were measured continuously on both lateral quartz balances. A home-made software, integrating systematic tooling factors and feeding back the quartz balance readings, was used to continuously tune the electrical power applied on both guns. Specular X-Ray Reflectivity measurements with a PANalytical  X'Pert MRD diffractometer were used to determine the thickness of evaporated single element layers and to calibrate the quartz balances of the evaporation chamber~\cite{Paratt_reflecto_1954}. Deposition was made at room temperature and at almost normal incidence and rotating the substrate continuously to ensure homogeneity and avoid specific atomic flux induced anisotropy~\cite{Hijikata_TbFe_uniformity_1990}. Different substrates (typically 10$\times$10~mm$^2$) have been tested, mainly Si$[100]$ with its native oxide,  SiO$\rm_x$(100~nm)/Si[100]. Even if not discussed here, successful growth was obtained as well on intrinsic and very resistive Si$[100]$ wafers and even Si$_3$N$_4$ amorphous membranes. After the ferrimagnetic layer growth, a capping layer was deposited also by evaporation. This could be Al or heavy metals like Pt or Ta, as discussed later on in Fig.~\ref{Tcomp_vs_compo_stacks}. 

\subsection{Rutherford Backscattering Spectroscopy~: depth-integrated composition}

  \begin{figure}[!h]
\includegraphics[width=7cm]{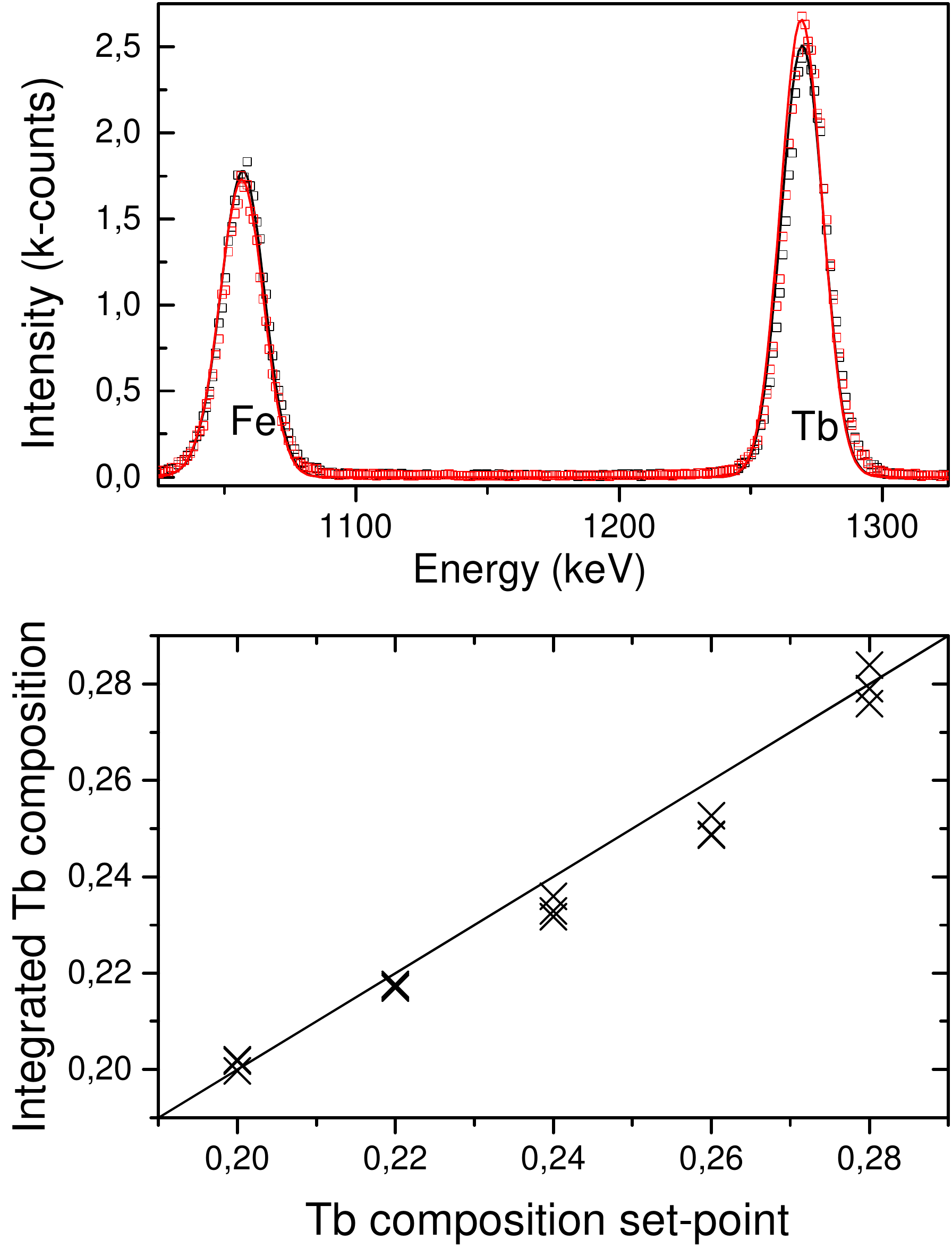}
\caption{a) RBS spectra on 7~nm Tb$\rm_{x}$Fe$\rm_{1-x}$ films: evaporation set points  $x=0.20$ (black) and  $x=0.22$ (red).  Fe and Tb peaks are indicated. The solid lines correspond to the adjustments of the raw data.  b) depth-integrated Tb composition in the Tb$\rm_x$Fe$\rm_{1-x}$ alloy as a function of the evaporation set-point. The different points correspond to successive measurements in different regions on the samples.   }
\label{RBS_TbFe}%
\end{figure}

The depth-integrated film composition  was determined by Rutherford Back scattering Spectroscopy (RBS). The incident beam of 2~MeV alpha particles with a maximum energy  spread of 100~eV was produced at the SAFIR (Paris) and SCALP platforms~\cite{Scalp_2017}. In both platforms, a 50~nA beam of diameter 0.5~mm was directed onto the samples at normal incidence. A Passivated Implanted Planar Silicon  detector of 2~mSr solid angle was placed at 165$^\circ$ to detect the He particles scattered from the sample. The scattered particle spectra which is proportional to the atomic number of the target atoms give the sample composition~\cite{RBS_book_1978}.  A set of 7~nm  thick Tb\rm$_x$Fe$\rm_{1-x}$ alloy films  with a Tb composition set-point ranging from 20 to 28~\% is discussed for illustration. In Fig~\ref{RBS_TbFe}a, RBS spectra of films of two different Tb concentrations are shown. The ratio of the Tb to Fe peak areas allows a direct measurement of the integrated composition in Tb$_x$Fe$\rm_{1-x}$. In the two films shown, $x$ is equal to  20 and 22~\% and the difference in the spectra is clearly visible to the eye, giving an indication of the sensitivity of the RBS measurement to the global film composition.  Fig~\ref{RBS_TbFe}b plots the effective depth-integrated Tb composition measured by RBS against the corresponding  set-point chosen for the evaporation process. The different points correspond to successive measurements in different area on the samples and a crude information on lateral homogeneity is here obtained. RBS results indicate a suitable integrated composition for the alloy and is used routinely to post-check the depth averaged Tb composition.  Note that RBS experiments can not be used in stacks including Ta or Pt because the peaks associated with those heavy elements are superimposed with that of Tb; the same holds for Co and Fe that can not be individually resolved. Therefore, only films covered with Al were measured with RBS. Below, integrated composition will be used to label the samples.  

\subsection{Scanning Transmission Electron Microscopy~: depth resolved analysis}\label{EELS}

\begin{figure}[!h]

a)\\
 \includegraphics[width=7.5cm]{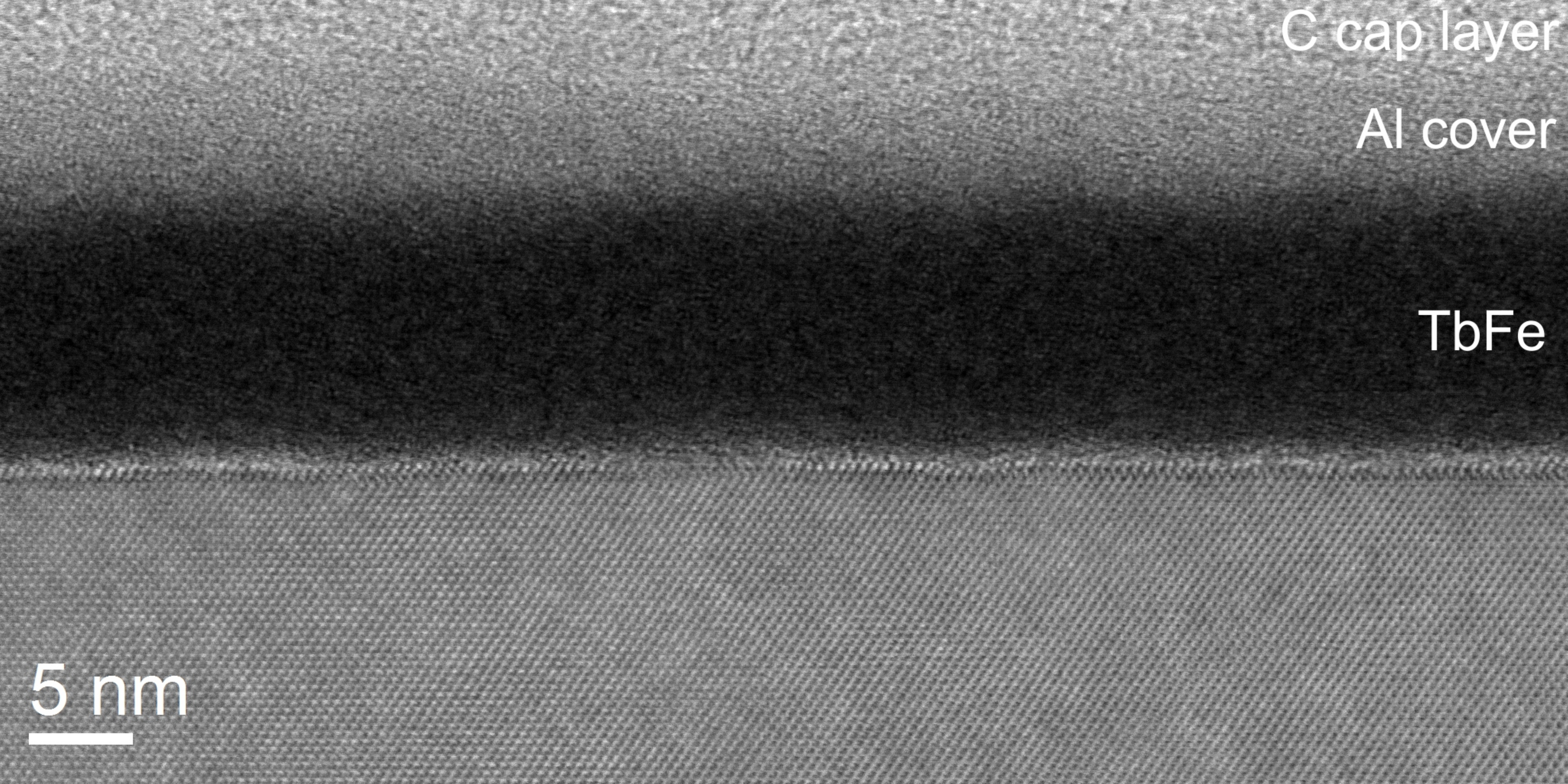}\\
\vspace{0.2cm}
b) \\
\includegraphics[width=7.5cm]{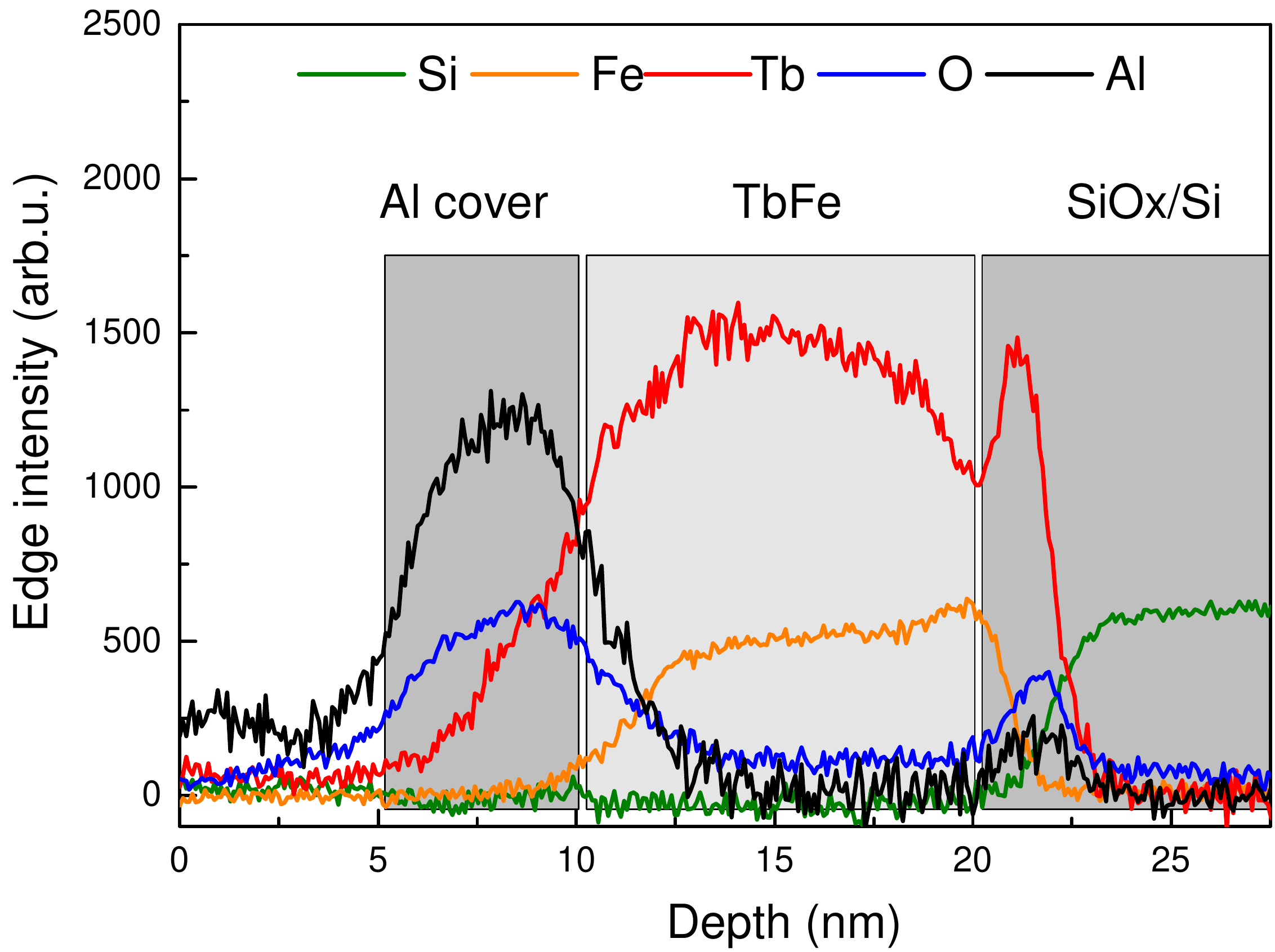}

\caption{a) STEM bright field image of a cross-section of a Si/ native Si0\rm$_x$/Tb$_{20.5}$Fe$_{79.5}$ 10~nm /Al 5~nm film. The amorphous structure of the constitutive TbFe layer is visible. b) EELS intensity profiles showing the depth location of the different element constitutive of the stack.}
\label{TEM_TbFe}%
\end{figure}

Scanning Transmission Electron Microscopy (STEM) was used to assess the crystallographic nature of the films and give information on the alloy top and bottom interfaces. For microstructural analysis, cross-sectional electron transparent samples were prepared by Focused Ion Beam on a SCIOS dual-beam platform following a standard procedure. To protect the sample surface during this preparation process, a thick carbon cap layer was deposited. 
STEM-EELS (Electron Energy Loss Spectroscopy) measurements were performed with an aberration corrected (Cs) STEM Nion UltraSTEM200 equipped with an in-house modified Gatan spectrometer~\cite{Kociak_2011}. The STEM was operated at 100 kV. A spectrum-line, in which a whole EELS spectrum is acquired for each probe position (0.1 nm apart), has been acquired across the film stack. In an EELS spectrum, the area under a characteristic edge is proportional to the number of analyzed atoms of the given chemical species per unit area. Such EELS elemental intensity profiles for the Al-L$_{2,3}$, Si-L$_{2,3}$, Tb-N$_{4,5}$, O-K and Fe-L$_{2,3}$ edges were measured. Absolute quantification requires information about different parameters among them the shell ionization cross section and is performed only in particular favorable situations. However in our case the calculation of the cross-section for Tb-N edge is subject to a large uncertainty. So we only used the signal intensity which is proportional to the amount of atoms present in the analyzed area~\cite{Kociak_2011}. An image of a 10~nm thick TbFe film is shown in Fig.~\ref{TEM_TbFe}a and EELS profiles are shown in Fig.~\ref{TEM_TbFe}b. The film thermal history (heated/cooled in air up to $\approx$ 120$^\circ$C) is typical of the baking temperature used for patterning wires investigated for Current Induced Domain Wall Motion (CIDWM). This temperature is also of the order of the film Curie temperature, reached during basic magnetic characterization.

The STEM image (Fig.~\ref{TEM_TbFe}a) shows that the alloy is fully amorphous and homogeneous in its bulk part, that the Al cover includes a few crystallized oxide particles and that the substrate is single crystalline with a very thin native oxide layer.  Chemical profiles across the film stack (Fig.~\ref{TEM_TbFe}b) further show that both top and bottom interfaces are perturbed. At the top, the TbFe/Al interface looks rough and/or mixed. EELS confirms that both Tb and Fe migrate into the Al cover and that terbium  extends further than iron into the aluminium cover. It also confirms the deep oxidation of the Al cover. At the bottom interface with the substrate, Tb and O intensities show superimposed and abrupt (about 2~nm) peaks at the location of the native SiO$_x$ layer covering the substrate.  This reflects both migration and oxidation of terbium of the bottom TbFe interface by the native silicon oxide with an enrichment in Terbium. We have integrated both EELS Tb and Fe intensities over the entire depth. We normalized the Tb and Fe ratio to 20.5\% (respectively 79.5\%) using of averaged proportion of Tb and Fe elements in the stack obtained by RBS measurements. The EELS profiles are not uniform in depth: the interfaces are richer in Tb depleting the bulk-like part. We obtain a negative bias of about 4\% in the composition of the bulk-like part of the alloy compared to the depth-integrated average composition.\\ 

In consequence, from the substrate to the top of the stack, the TbFe alloy concentration profile is i) at the bottom interface, an oxidized layer with a Tb excess followed by a thin layer with a Tb deficit (up to 30\% less) ii) a bulk-like TbFe alloy film with an almost constant effective composition of 16.5\% (a bias of about -4\% from the total RBS composition of 20.5\%) and iii) a top interface slightly mixed with the Al cover, also lacking a bit of terbium sunk by the aluminium oxide.  The general features are in agreement with former results on GdCo films~\cite{Bergeard_PRB_2017_GdCo} showing Gd surfactant and oxidation effects. This also supports a thickness dependance of the magnetic properties of the samples, as demonstrated in the following but in old and recent papers as well~\cite{Dover_1986,Blanchard_1990,Cid_2017,Mangin_PRB_2016_thickness_dependance,Albrecht_FM_2016_TbFe_thickness_dependance,Ono_APEX2015_GdFeCo_Chiral_DW}. We next discuss how this bias  in the effective composition impacts on magnetic properties. 

\subsection{Magnetization reversal and compensation points in TbFe (TbFeCo)}

\subsubsection{Above room temperature, from magneto-optical investigations} 

\begin{figure}[!h]
\includegraphics[width=6cm]{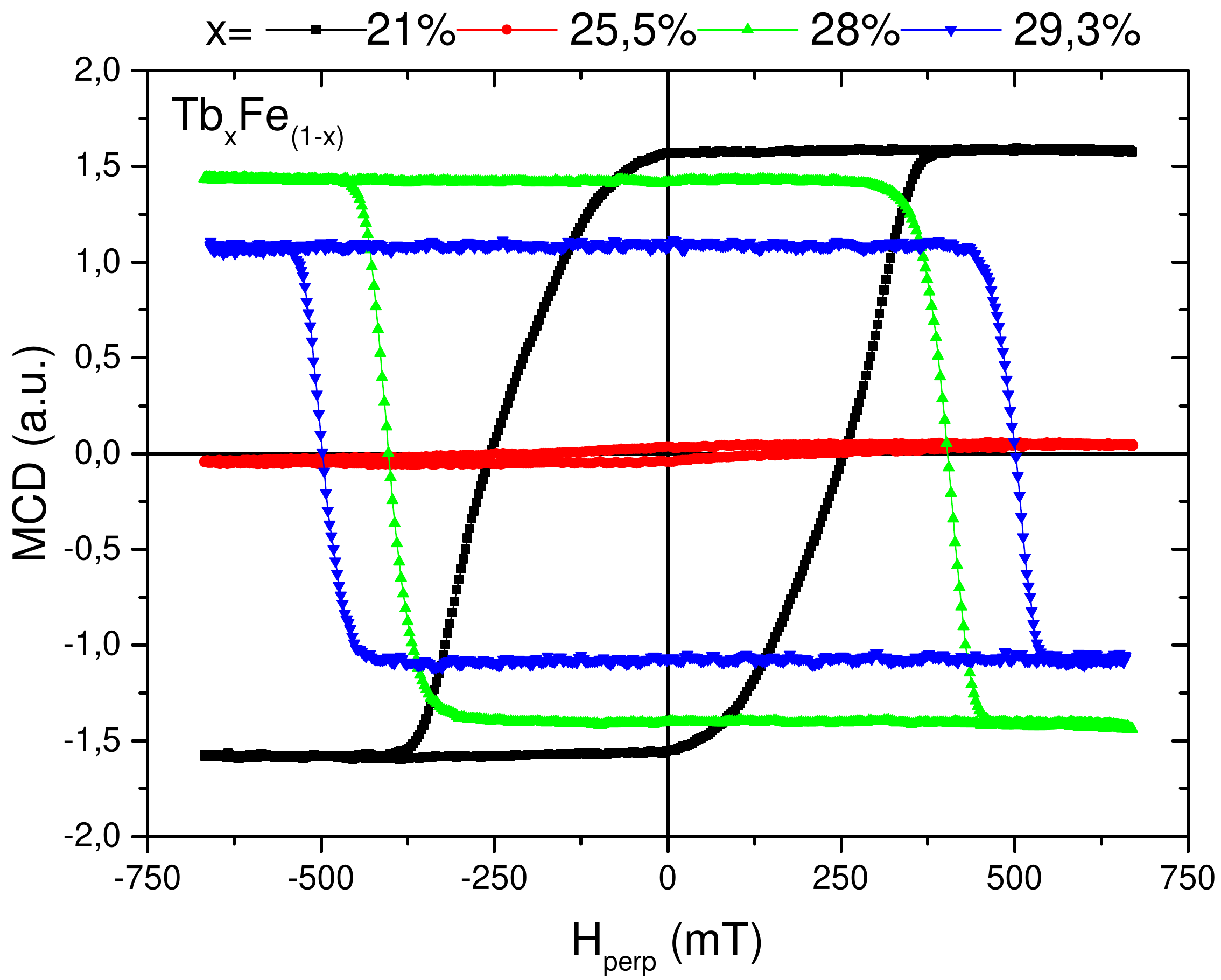}
\includegraphics[width=6cm]{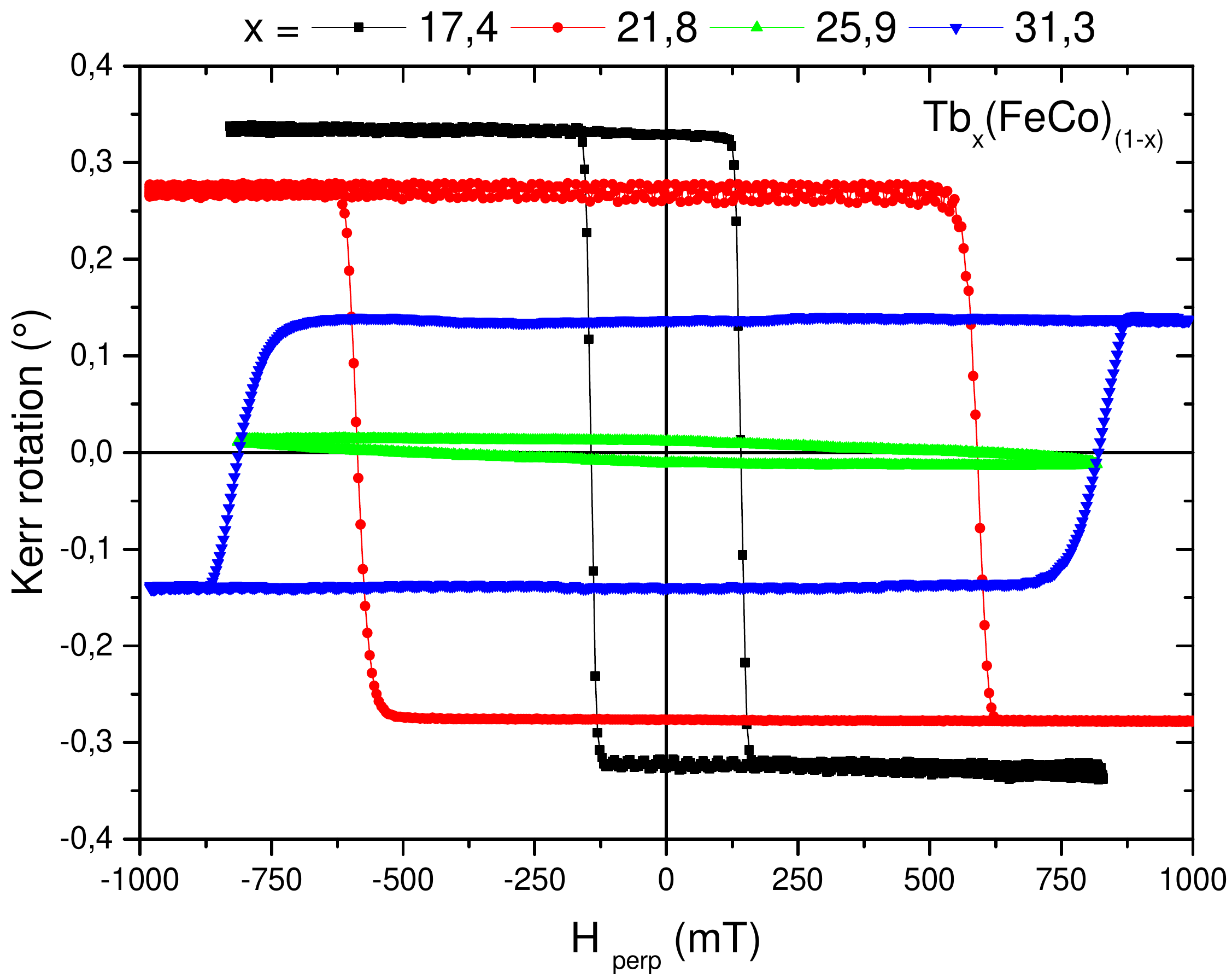}
\caption{Polar magneto-optical hysteresis loops of 7~nm a) Tb\rm$_{x}$Fe$\rm_{1-x}$ (Magnetic Circular Dichroism) and b) Tb\rm$_{x}$FeCo$\rm_{1-x}$ (Kerr rotation)  films for $x$ ranging from 0.2 to 0.3 at room temperature. Both sets are grown on Si/SiO\rm$x$ and 5~nm Al covered}
\label{TbFe_TbFeCo_MOKE_RT_vs_compo}%
\end{figure}

\begin{figure}[!h]
\includegraphics[width=7.5cm]{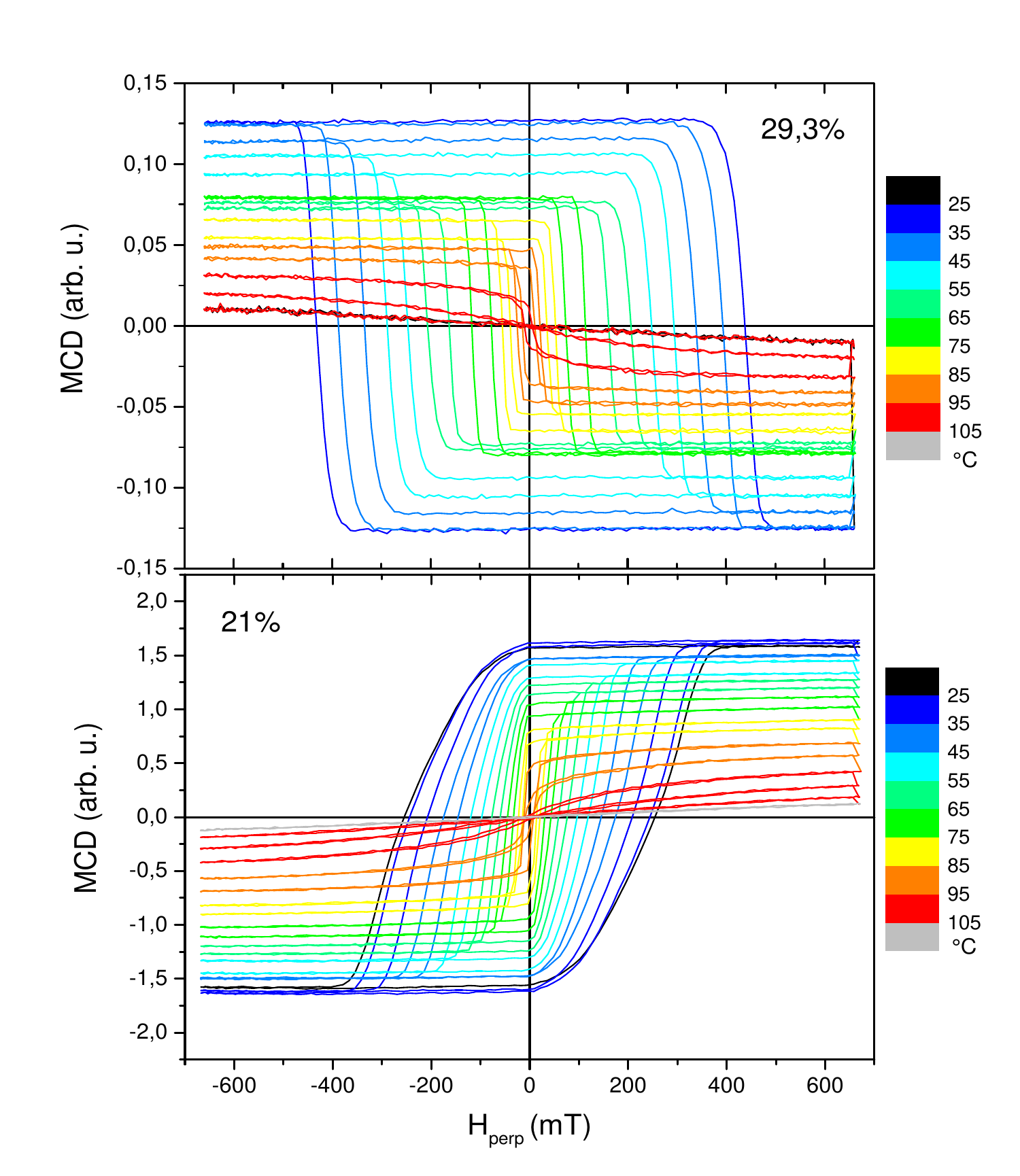}
\caption{Polar magneto-optical hysteresis loops of 7~nm thick  Tb$_{21}$Fe$_{79}$ and Tb$_{29.3}$Fe$_{69.7}$ films as function of temperature in $^\circ$C, indicated by the color bar on the figure.}
\label{TbFe_MOKE_temperature}%
\end{figure}
The macroscopic magneto-optical hysteresis loops of TbFe/TbFeCo alloys films were obtained  in the polar geometry, i.e., at quasi normal incidence with a magnetic field perpendicular the sample plane to check the out-of-plane magnetization component.  A red $\lambda$=652~nm (1.85~eV) laser was used and for such a wavelength, optical process occurs only for the 5d RE and 3d TM conduction electrons. The 3d TM moments being much larger, either Kerr rotation or Magnetic Circular Dichroism (MCD) are mostly only proportional to Fe (FeCo) magnetization in our films. For magneto-optical reasons, the   MCD and Kerr rotation show opposite signs~\cite{hansen-JAP-66-756-1989,Rasing_PRL_MO_TbFe_element_specific_2013}. As $x$ increases and the magnetic compensation is crossed, the dominant magnetic sub-lattice that is coupled parallel to an external magnetic field is changed so that a Kerr (or MCD) reversed loop is expected from each side of the compensation~\cite{Ferre_JMMM_94_TbFe}. 

Magneto-optical hysteresis loops of 7~nm Tb\rm$_{x}$Fe$\rm_{1-x}$  films (respectively Tb\rm$_{x}$(FeCo)$\rm_{1-x}$) are shown in Figure~\ref{TbFe_TbFeCo_MOKE_RT_vs_compo}a (resp.~b). In both set of films, compensation can be identified by the reversal of the hysteresis loops and the very flat hysteresis loops that illustrate the divergence of the coercivity in the vicinity of compensation. The compensation temperature of the Tb poor film (black) is below room temperature, the intermediate Tb composition (red/blue) corresponds to compensation around room temperature whereas the Tb rich films (green/blue) have a compensation temperature above room temperature.  In TbFe films, compensation around room temperature is obtained for about 27\% Tb whereas in TbFeCo films, it's about for 28\% Tb. A  shift of ~1\% Tb is measured to get room temperature compensation when comparing TbFe and TbFeCo; this may be understood knowing that oxidation rate of TbFe is hindered by cobalt which slightly shifts compensation~\cite{Miller1988}. The main result is the decrease of the effective compensation composition as compared to bulk~\cite{hansen-JAP-66-756-1989}.

\begin{figure}[!h]
\begin{center}
\includegraphics[width=0.5\columnwidth]{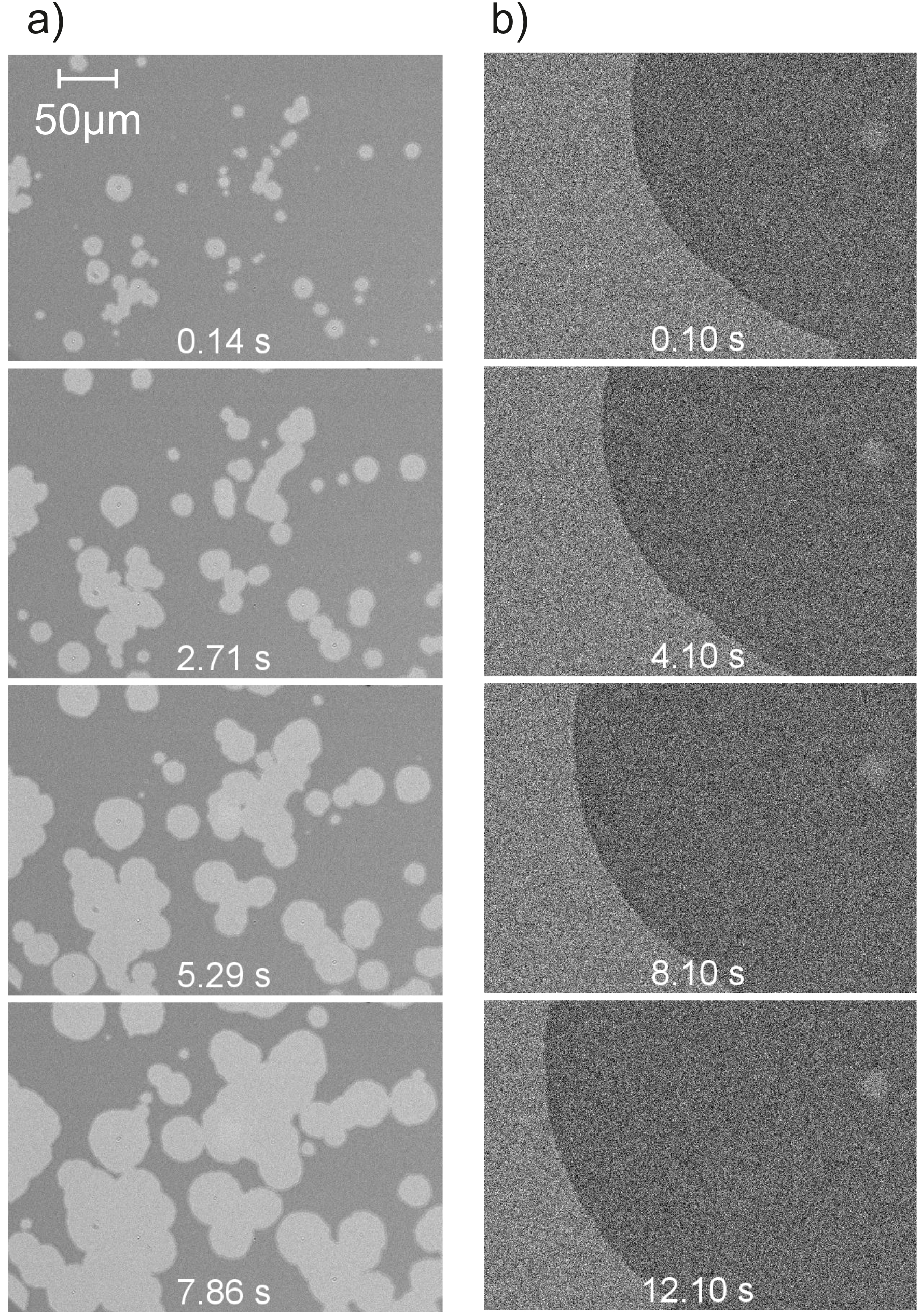}
\end{center}
\caption{Magneto-optical images illustrating DW nucleation and propagation in 7~nm thick films a) Tb$_{24}$Fe$_{76}$ (B=430~mT) and b) Tb$_{28}$Fe$_{72}$  (B=600~mT) at room temperature.  }
\label{MOKE_reversal}%
\end{figure}

In TbFe films, compensation above RT (Tb-rich case) is difficult to observe because $\rm T_{comp}$ is close or equal to the Curie temperature. Above RT, a selection of illustrative hysteresis loops of two 7~nm Tb\rm$_{x}$Fe$\rm_{1-x}$  films from either side of compensation is shown as a function of temperature in Figure~\ref{TbFe_MOKE_temperature}. As expected, both films exhibit loops reversed one respective to the other. The Curie temperature is around 100$^\circ$C for both films. The anisotropy of films remains out-of plane and the loops keep their remanence up to high temperature for both films. A significant difference however appears: the Tb-rich films exhibit very square loops which is usually associated with rare nucleation events at a given magnetic field followed by rapid domain wall propagation. The hysteresis loops of the films that are Fe dominated above room temperature are always rounder. We selected 7~nm thick Tb$_{28}$Fe$_{72}$ (Tb rich) and Tb$_{24}$Fe$_{76}$ (Tb poor) films similar to those shown in Figure~\ref{TbFe_MOKE_temperature} and investigated the magnetization reversal process by Kerr microscopy~\cite{McCord_review_microscopy_2015}. Reversal occurs by domain wall nucleation and propagation in the two films and outlines the great quality of the films, with a very few amount of extrinsic defects. As guessed from the hysteresis loops, Figure~\ref{MOKE_reversal} illustrates the significant difference in nucleation. This is agreement with former results showing a difference in difference stability, domain size and magnetization reversal above and below compensation, evidenced in the nucleation field and its dispersion~\cite{Mochida2002,Gadetsky1992,Takahashi1986}. 

\subsubsection{From low temperature and up to large fields using Extraordinary Hall Effect,  VSM and AGFM}

\begin{figure}[!h]
\includegraphics[width=0.7\columnwidth]{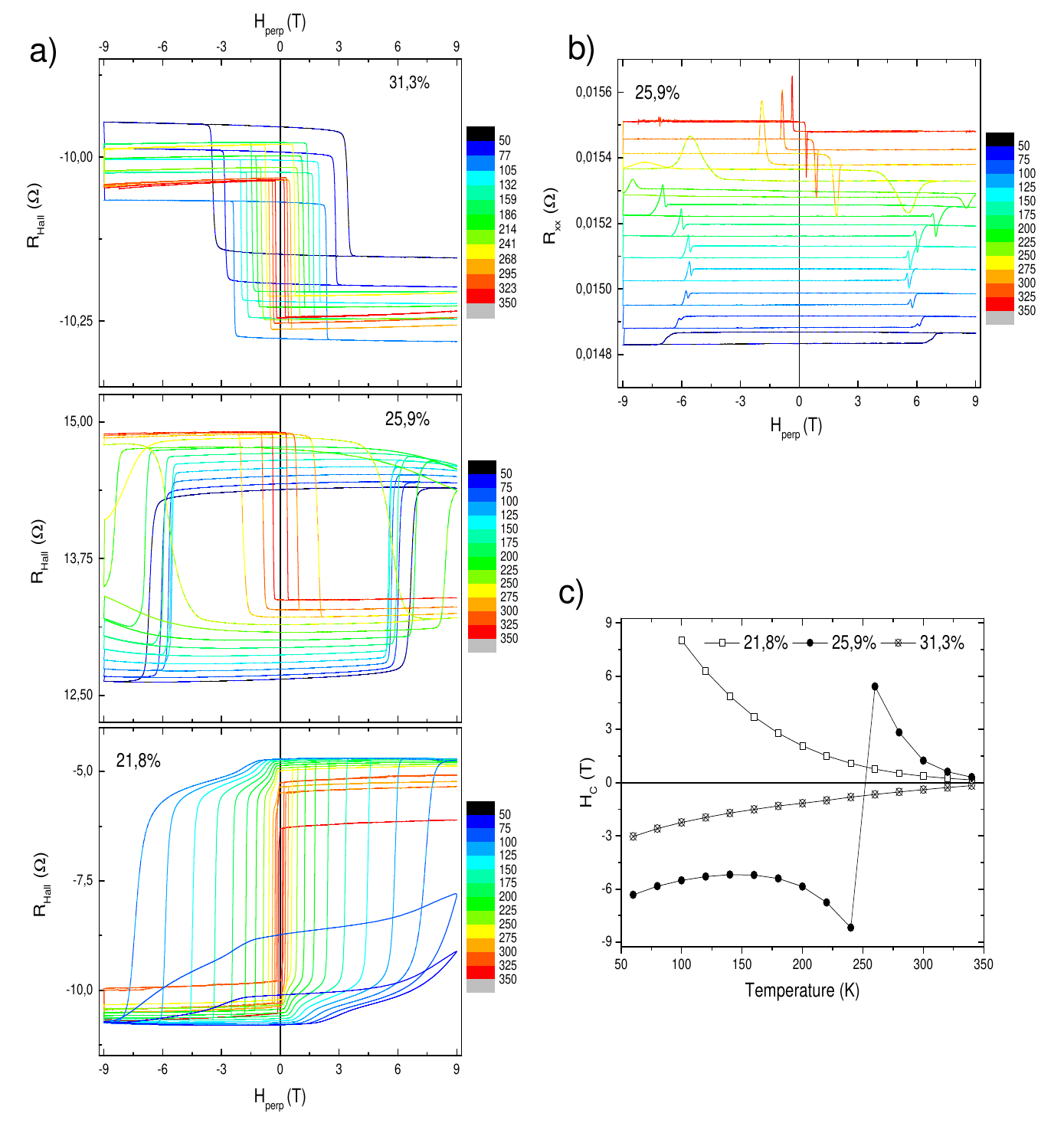}
\caption{Transport experiments results on films of 7~nm Tb\rm$_{x}$FeCo$\rm_{1-x}$ films for $x$ ranging from 0.218 to 0.313 grown on Si/100~nm SiO$x$ and 5~nm Al covered. a) EHE hysteresis loops from T=60~K to 350~K as indicated by the vertical scale.  b) $R_{xx}$ loops for the 0.259 film. c) Coercive fields $H_C$ as a function of temperature.}
\label{TbFeCo_EHE_vs_compo_loops}
\end{figure}

The samples were further investigated using a commercial Quantum Design Pluri Physical Measurement Systems (PPMS) that enables to measure both transport properties (Extraordinary Hall Effect EHE) and magnetization by Vibrating Sample Magnetometer (VSM), with magnetic field (-9T/+9T) applied perpendicular to films in a temperature range between 2~K and 400~K.  EHE and magnetoresistance measurements were simultaneously performed employing the four-probe technique. Since TbFeCo resistivity is very large~\cite{Gimenez1988}, shortcuts can be easily obtained with any cracks or with a unsuitable substrate: films must be grown on an insulating substrate e.g. a 0.1$\mu$m SiO\rm$_x$ layer is required to prevent current flowing into the substrate. The set of TbFeCo films of Fig.~\ref{TbFe_TbFeCo_MOKE_RT_vs_compo}b is used to illustrate the EHE results on full films.  Hysteresis loops are shown in Fig.~\ref{TbFeCo_EHE_vs_compo_loops}. Like in magneto-optics, the EHE signal comes from conduction electrons: the Transition Metal dominates and the Rare Earth has nevertheless a tiny contribution of opposite sign~\cite{Malmhall_EHE_TbFe_values_1983,Mansuripur_EHE_TbFe_1991,Malmhall_TbFe_vs_T_1985}. Altogether, the sign of the Hall signals is  meaningful and EHE loops reverse when the ferrimagnet goes through compensation. When the current flows along $x$ the magnetic field along z, the Hall electric field is along y. Here, the Hall resistance $R_H=R_{xy}$ is positive when the TM sublattice is parallel to the external field. The thermal and composition evolution of the loops provide further evidence of a FeCo (Tb) dominated behavior for low (large) Tb  amount and the existence of compensation for the intermediate composition. This is further demonstrated in Fig.~\ref{TbFeCo_EHE_vs_compo_loops}c gathering the thermal evolution of the coercive field for the 3 typical samples. Note that VSM hysteresis loops described with the same magnetic field sweeping rate show exactly the same coercive fields. $R_{xx}$ loops were measured simultaneously with the $R_{xy}$ Hall resistance. Loops  are shown  for the film of intermediate composition in Fig.~\ref{TbFeCo_EHE_vs_compo_loops}b.  They show  peaks at the exact values of the coercive fields as observed previously~\cite{Yumoto1988,Cheng2005}. Those are antisymmetric and understood as resulting from DW perpendicular to the magnetization and to the current during the reversal~\cite{Cheng2005}. This does not depend on the position relative to compensation. Another point that we observe is the $R_{xx}$ decrease as T is increased; this is typical of amorphous metals as already discussed in reference~\onlinecite{Fert1979}. \\

Neither EHE  nor  MOKE amplitude enable quantitative magnetization measurements. In order to obtain magnetization values representative of the thermal evolution of the difference between the 3d TM parallel to 5d RE moments and the antiparallel 4f RE moments, hysteresis loops were measured with VSM and with a commercial Alternating Gradient Force Magnetometer (AGFM) from Princeton Measurement Corporation allowing working at room temperature with magnetic field up to 1~T. The typical sensitivity of VSM with such configuration is around 10$^{-6}$ emu. The AGFM system has about the same limit of sensitivity. For very thin films like ours of typical volume around 10$^{-8}$cm$^3$, the two set-ups lead to a magnetization sensitivity of about 100 emu/cm$^3$ = 125~mT. Extracting the signal arising from the ferrimagnetic layer in question from the background signals requires: a) minimization of the sample holders magnetic signature and b) a means for subtracting substrate contributions. With both techniques, apart from a large diamagnetic contribution relatively easy to remove, a S shape contribution was evidenced around H=0. This contribution, temperature independent in our investigation range,  was attributed to the substrate. Its removal at low temperature where the ferrimagnetic film has a huge coercivity was quite easy but sometimes a bit more tricky at larger temperature.

\begin{figure}[!h]
\includegraphics[width=9cm]{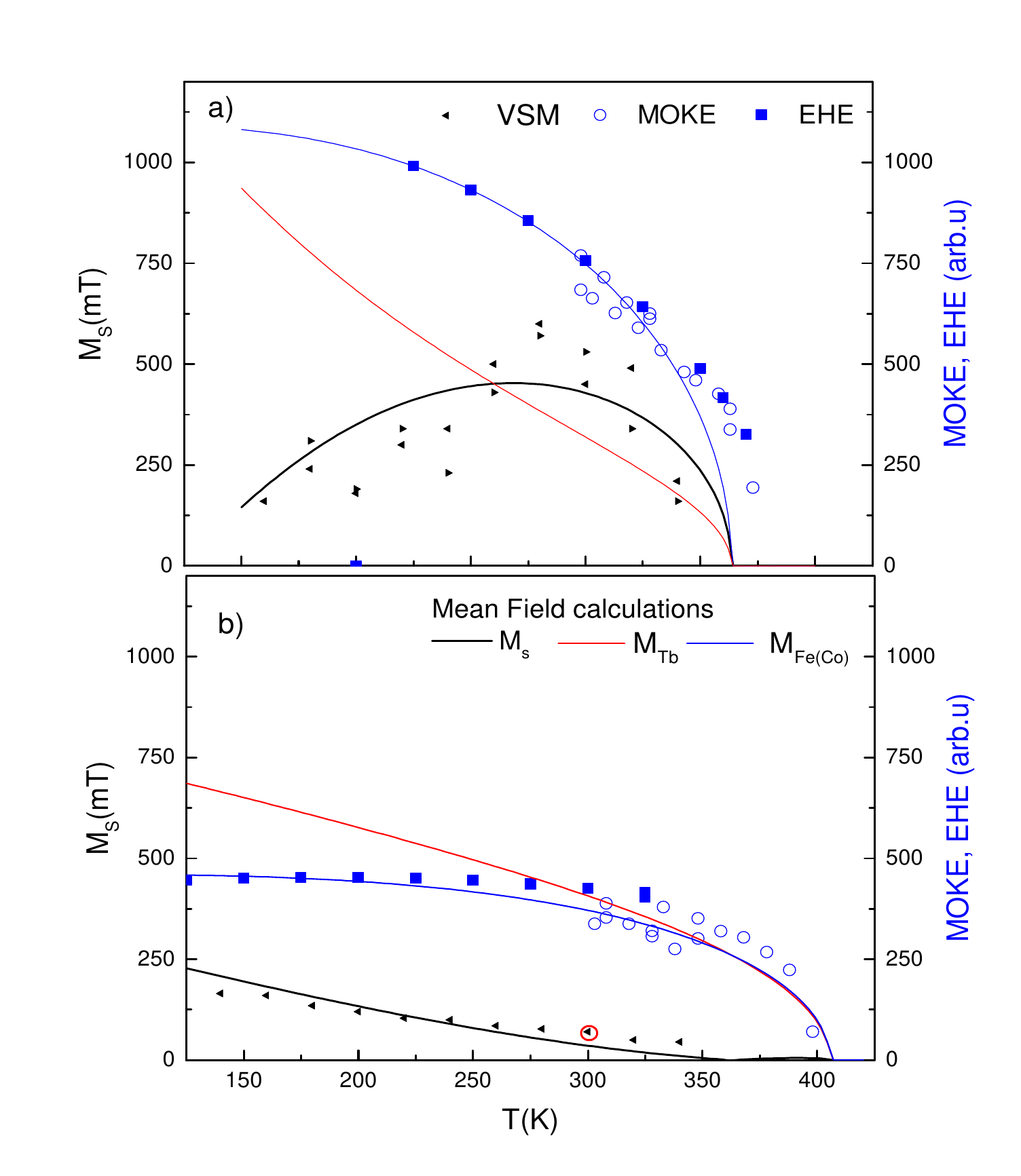}
\caption{Magnetization versus temperature  in Si/100nm SiO\rm$x$/7nm Tb$_{0.21}$Fe$_{0.79}$/5nm Al (a) and  Si/100nm SiO\rm$x$/7nm Tb$_{0.31}$FeCo$_{0.69}$/5nm Al (b) films from VSM (triangles) and AGFM (single red circle). MOKE (above room temperature - dots) and EHE (squares) loops amplitude have been superimposed and are mainly proportional to the polarization P that reflects the TM magnetization. Mean field calculation results as described in~\cite{hansen-JAP-66-756-1989} are shown with solid lines.}
\label{TbFeCo_VSM_all_combined}%
\end{figure}

Figure~\ref{TbFeCo_VSM_all_combined} shows magnetization as a function of temperature for a TbFe film (a) (Fe dominated, $\rm T_{comp}$ below 150~K) and its counterpart, a  TbFeCo film (b) (Tb dominated,  $\rm T_{comp}$ above 400~K).  Close to compensation,  some of the coercive fields were so large that we could not measure full hysteresis by VSM. The solid lines are the results of mean field calculations, as detailed in reference~\onlinecite{hansen-JAP-66-756-1989}, that have been superimposed to the experimental results for the magnetization. They also provide the individual sublattice magnetization and their different thermal evolution. MOKE amplitudes obtained above room temperature are shown on the same graph.  MOKE data  mainly given by the large 3d TM and tiny 5d RE contributions are essentially proportional to the spin polarization and can merge onto the EHE data obtained on wider temperature range. Both adjust reasonably to the TM sublattice magnetization obtained from the mean field calculations. Within the limits of sensitivity of our techniques, all results are consistent and show that our TbFeCo have indeed a small magnetization above room temperature but keeps a significant polarization that can be easily probed by  MOKE or EHE.

\subsection{Magnetic properties depending on the stack}

\subsubsection{As a function of the thickness, in TbFe alloys}

\begin{figure}[!h]
\includegraphics[width=7cm]{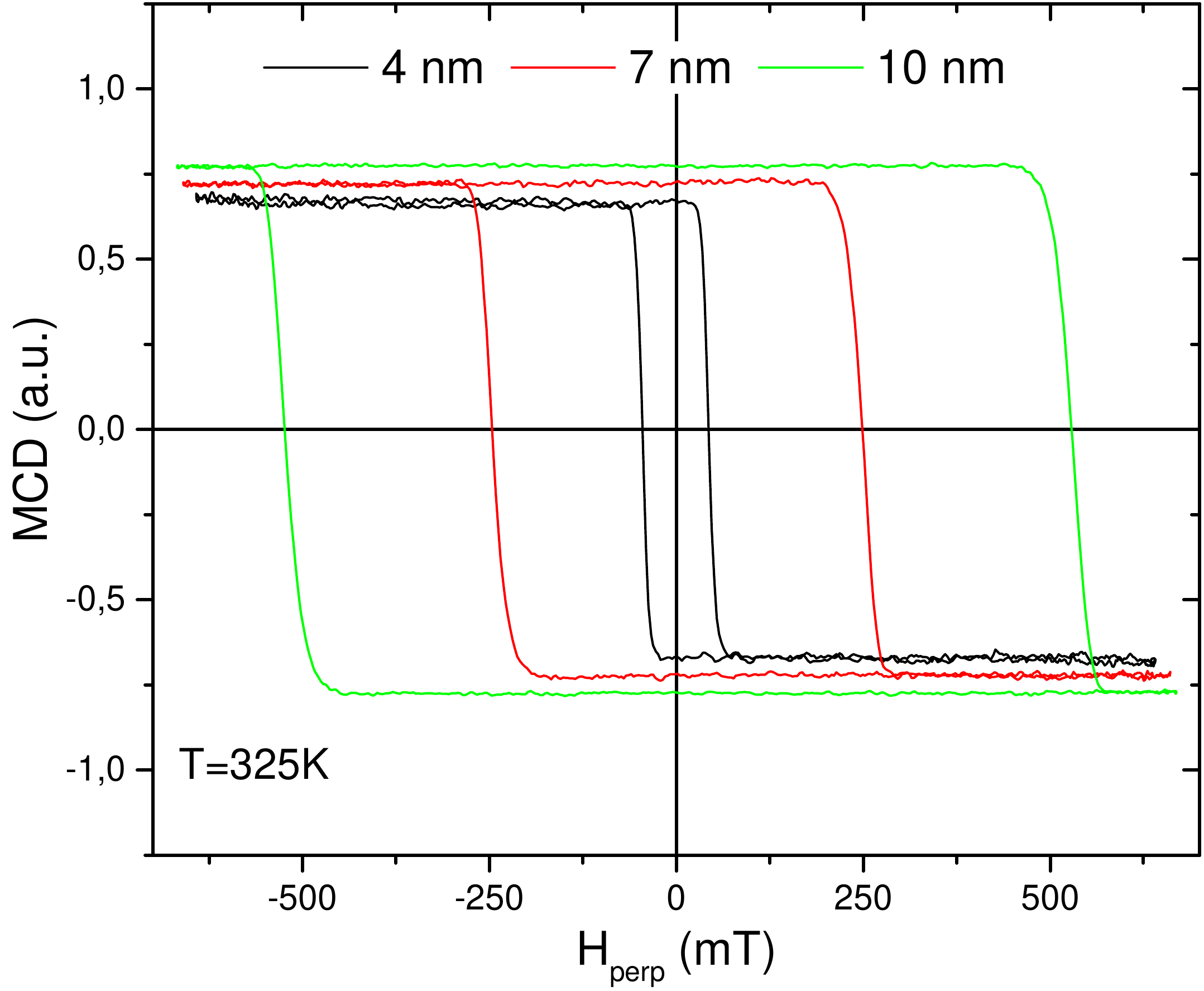}
\caption{Polar hysteresis loops of Tb$_{28}$Fe$_{72}$ films of different thickness (indicated in the figure) at T=325K.  }
\label{G_MOKE_thickness}%
\end{figure}

Thin films of decreasing thickness and fixed composition were grown.  MOKE (Figure~\ref{G_MOKE_thickness}) and EHE (not shown) experiments demonstrate that the coercivity increases as the thickness of the alloy film is increased from 4 to 10~nm. Experimentally, the large coercivity of the films close to compensation is difficult to overcome and it was not possible to saturate the 10~nm thick films in our MOKE set-up at room temperature; that's why the loops are shown at 325~K. Additionally, when increasing the thickness, the compensation temperature is shifted towards larger values. This is in agreement with previous results describing a similar thickness dependence of coercivity and even polarity of the hysteresis loop of films prepared in the same growth conditions~\cite{Malmhall_EHE_versus_t_1982,Mangin_PRB_2016_thickness_dependance}. They were interpreted as indicating  a change in the effective film composition revealing the importance of surface vs bulk~\cite{Scheinfein1991}. This is here supported by the Tb\rm$_{x}$Fe$\rm_{1-x}$ alloy films EELS analysis (see section~\ref{EELS}). 
Firstly, a part of the terbium atoms present in the films is not available for its bulk part and for small thicknesses, a 28\% film may behave like its bulk region 24\% one at the same temperature i.e. with an apparent smaller $\rm T_{comp}$ than its thicker counterpart. Secondly, interfacial regions where the local composition is quite different can impact significantly the nucleation field and consequently coercivity. Finally, it's well known that nucleation is affected by defects whose density is increased as the thickness is reduced. 

\subsubsection{As a function of the buffer/cover, in TbFeCo alloys}

\begin{figure}[!h]
\includegraphics[width=9cm]{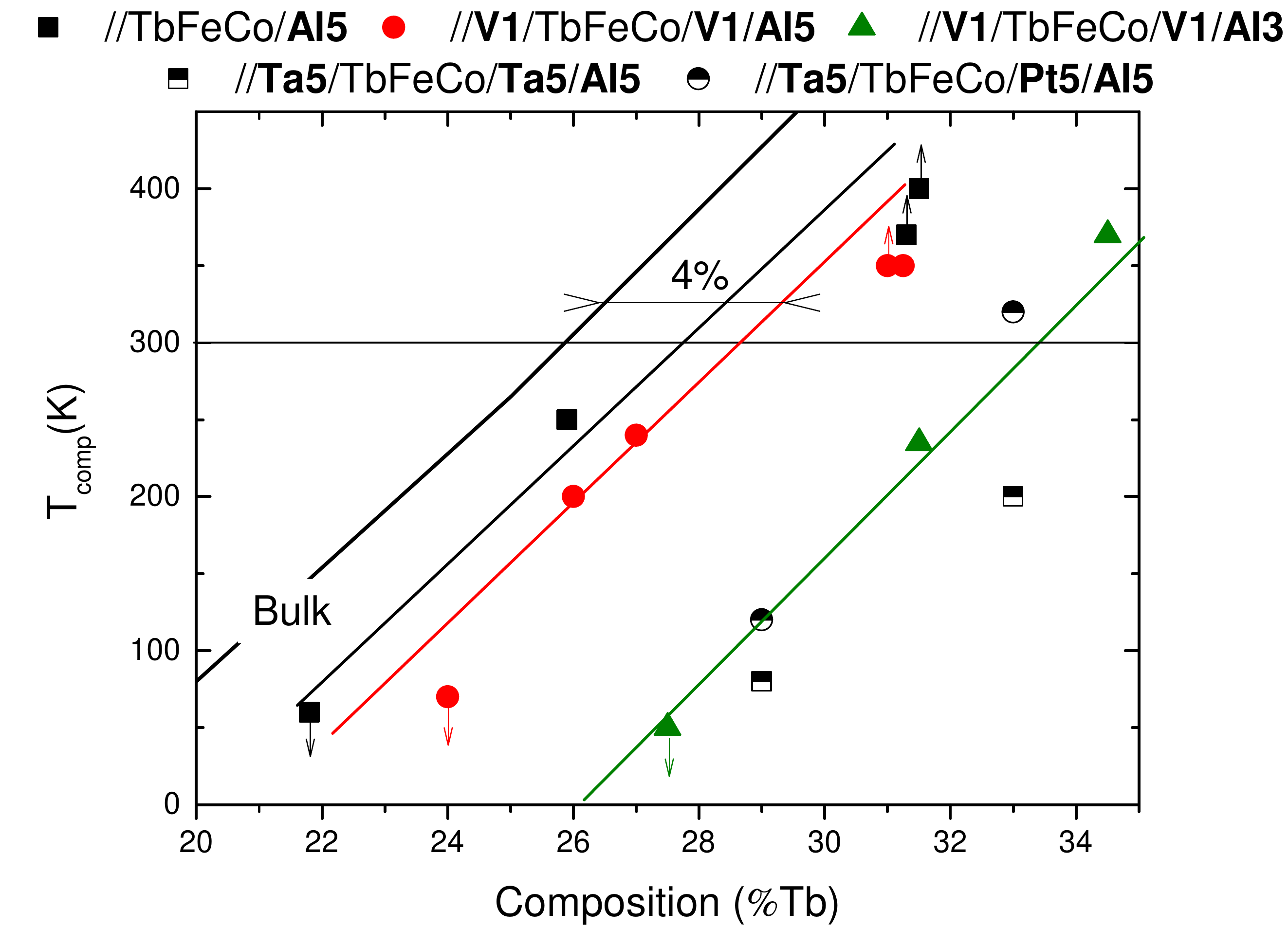}
\caption{Evolution of the $\rm T_{comp}$  in different Tb\rm$_{x}$Fe$\rm_{1-x}$ 7~nm stacks versus the Tb composition. Different symbols are used. When arrows are drawn, it means that the samples did not transit up to this temperature in our experimental conditions and that the compensation if it exists, is above (up arrow) or below (down arrow) the point.  The thick black solid line corresponds to the bulk evolution of $\rm T_{comp}$ reported in~\cite{hansen-JAP-66-756-1989}. Other solid lines shown for films embedded V and Al are supported by the assumption that $\rm T_{comp}$ is linear with the Tb amount in the investigated range, like in the bulk, with a slope of about 40~K per percent. For Pt and Ta buffer/cover layers, significant error bars may exist for the composition. }
\label{Tcomp_vs_compo_stacks}%
\end{figure}

To further investigate the diffusion processes and the preferential oxidation of the rare earth that occurs at the interfaces,  simple films of fixed thickness have been encapsulated with vanadium and aluminium or embedded between layers of heavy metals like Ta and Pt. The evolution of the $\rm T_{comp}$ in different 7~nm Tb\rm$_{x}$FeCo$\rm_{1-x}$ stacks is presented in Figure~\ref{Tcomp_vs_compo_stacks} versus the composition. 

The thick black solid line corresponds to the bulk evolution as given in reference~\onlinecite{hansen-JAP-66-756-1989}. The smaller the terbium amount, the smaller  $\rm T_{comp}$ that shifts towards low temperature at a rate of about 40~K per Tb percent. We repeated all steps of basic characterization in the different film stacks. Let us comment first on films encapsulated with vanadium but covered with 5~nm of aluminium to be compared with the previous set of samples. A clear shift towards low $\rm T_{comp}$ is observed in comparison with bulk like alloys. It further reveals a 4\% biased effective composition in thin films as obtained from RBS and EELS experiments. We tried vanadium with the idea that it could prevent diffusion like niobium does efficiently for TbFe~\cite{PRB_Nb_hansen} and because of its small spin-orbit (the later being an important parameter for ongoing CIDWM studies).  However, the compensation of stacks including or not the 1~nm vanadium layer is not significantly different and corresponds maximum to a 1\% reduction of Tb composition.   The effect of the biasing in the effective composition is even more drastic in the second type of stacks. When the Al cover thickness is reduced to 3~nm, the compensation is shifted to much lower temperature and shows that a much larger amount of terbium migrated towards the top and oxidizied cover layer, as shown in TbFeCo depending on the Al buffer quality~\cite{Ohta_2003}.  It reveals that terbium migrates across vanadanium that does not act as an efficient diffusion barrier. For ferrimagnetic films embedded between heavy metals like platinum or tantalum, the alloy composition may exhibit a significant error since RBS experiments are meaningless. Therefore, no solid lines have been drawn but the tendency is obvious: a similar bias of composition occurs.   This again reveals the very large sensitivity of bulk compensation to the amount of terbium sunk in the interfaces. 

\section{Conclusion}

In this paper, Tb\rm$_x$Fe$\rm_{1-x}$ or Tb\rm$_x$FeCo$\rm_{1-x}$ alloys thin films grown by co-evaporation were investigated by combining structural, transport and magnetic investigations. The films exhibit the desired integrated Tb composition and are amorphous. 
We also evidenced a non homogeneous depth distribution of the Rare Earth within the film depth. Terbium, being very sensitive to its local environment, can be sunk by buffer/cover layers, in particular if they contain oxygen. This leads to Tb oxidizied interfaces and a biased magnetic compensation temperature that reflects the reduced Tb amount of the bulk of the layer. We also evidenced that this shift in magnetic properties may be modified by changing the nature or thickness of the buffer/cover layers. \\

A very appealing property, when thinking about experiments that require to inject a significant amount of current in the wire, is that the anisotropy remains out-of plane and the loops keep their remanence up to high temperature, as required for further CIDWM experiments. Moreover, magnetization reversal occurs by rare nucleation and a smooth domain wall propagation. \\

Already in such a simple TbFe stack, it's very interesting to note the asymmetric character of the considered interfaces, crucial for properties like Dzyaloshinskii-Moriya interaction~\cite{Thiaville_EPL_2012,Ono_APEX2015_GdFeCo_Chiral_DW}. Any further studies of domain wall propagation in such systems must take into account the detailed structure of the stacks.

\begin{acknowledgments}
A. Pointillon would like to thank the R\'egion Ile-de-France (C'NANO IdF ``FAST'' project). O. Rousseau was granted by the project MULTIDOLLS from the French National Research Agency (ANR) ANR-12-BS04-0010-02. The authors thanks S. Rohart for his constant implication in the MBE development. The authors acknowledge for VSM measurements financial supports from R\'egion  Ile de France  (C'Nano IdF ``NOVATECS'' Project No IF-08-1453.R) and from the French National Research Agency (ANR) as part of the ``Investissements d'Avenir'' program (Labex charmmmat,  ANR-11-LABX-0039-grant). 
\end{acknowledgments}

\bibliography{TbFe_biblio}

\end{document}